\documentclass[aps,prl,one,superscriptaddress]{revtex4}

\usepackage{amssymb}
\usepackage{amsfonts}
\usepackage{rotating}
\usepackage{amsmath}
\usepackage{graphics}
\usepackage{palatino}
\usepackage{epsfig}
\usepackage[pdftex,usenames]{color}
\usepackage{hyperref}

\newcommand{\be}{\begin{equation}}
\newcommand{\ee}{\end{equation}}
\newcommand{\av}[1]{\langle #1 \rangle}
\newcommand{\bea}{\begin{eqnarray}}
\newcommand{\eea}{\end{eqnarray}}

\begin{document}


\title{Controlling Contagion Processes  in Time-Varying Networks}

\author{Suyu Liu}
\affiliation{State Key Lab. of Industrial Control Technology, Institute of Cyber-systems and Control, Zhejiang University, Hangzhou 310027, China}
\affiliation{Laboratory for the Modeling of Biological and Socio-technical Systems, Northeastern University, Boston MA 02115 USA}

\author{Nicola Perra}
\affiliation{Laboratory for the Modeling of Biological and Socio-technical Systems, Northeastern University, Boston MA 02115 USA}

\author{M\'arton Karsai}
\affiliation{Laboratory for the Modeling of Biological and Socio-technical Systems, Northeastern University, Boston MA 02115 USA}

\author{Alessandro Vespignani}
\affiliation{Laboratory for the Modeling of Biological and Socio-technical Systems, Northeastern University, Boston MA 02115 USA}
\affiliation{Institute for Scientific Interchange Foundation, Turin 10133, Italy}

\date{\today} \widetext

\begin{abstract} 
The vast majority of strategies aimed at controlling contagion processes on networks considers the connectivity pattern of the system as either quenched or annealed. However, in the real world many networks are highly dynamical and evolve in time concurrently to the contagion process. Here, we derive an analytical framework for the study of control strategies specifically devised for time-varying networks. We consider the removal/immunization of individual nodes according the their activity in the network and develop a block variable mean-field approach that allows the derivation of the equations describing the evolution of the contagion process concurrently to the network dynamic.  We derive the critical immunization threshold and assess the effectiveness of the control strategies. Finally, we validate the theoretical picture by simulating numerically the information spreading process and control strategies in both synthetic networks and a large-scale, real-world mobile telephone call dataset. 
\end{abstract}

\maketitle

\begin{figure}
\includegraphics[width=10cm]{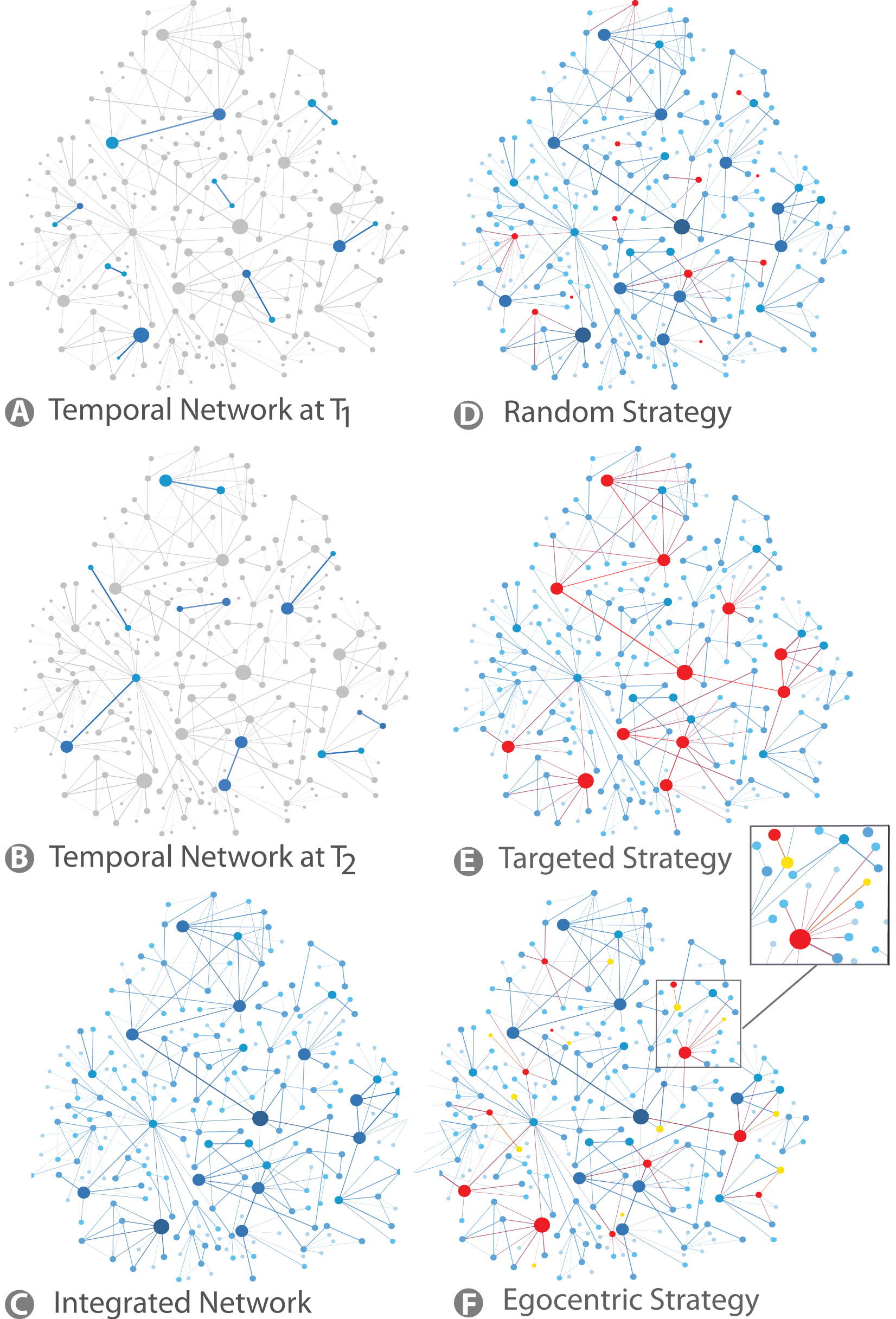}
\caption{{Schematic representation of Activity driven model and control strategies. Panels A and B show the temporal network at two different time steps $T_1$ and $T_2$. Panel C presents the integrated network over a certain period of time, in which node size and color describe the activity, while link width and color represent the weight. Panels D, E and F show random, targeted and egocentric control strategy respectively. Node color is assigned as its state, immunized as red and probes as yellow.}}
\label{fig:Fig1}
\end{figure}

\begin{figure}
\includegraphics[width=10cm]{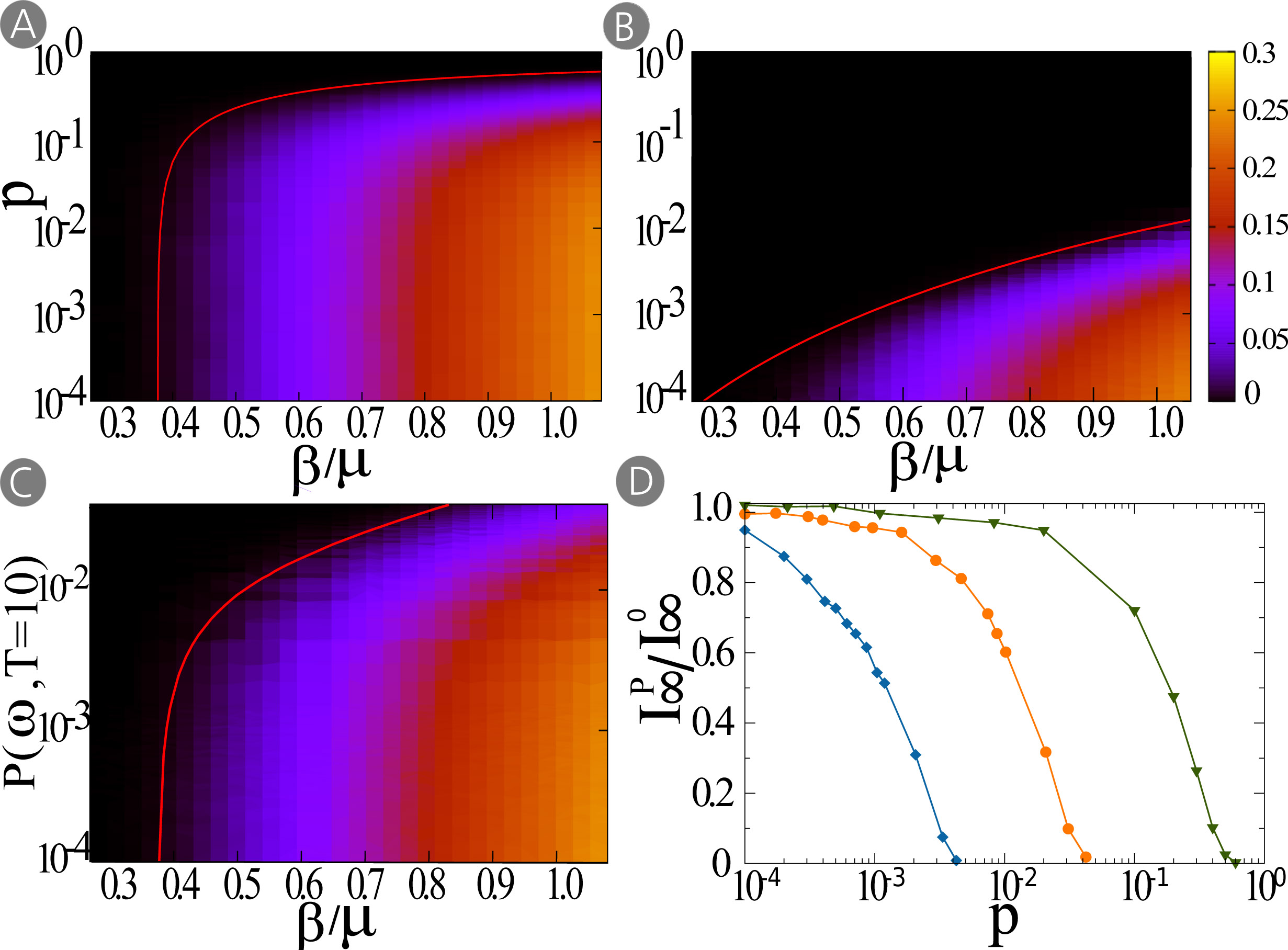} 
\caption{Panels A, B and C show the phase space of an SIS process under random, targeted and egocentric control strategy respectively. Considering $N=10^4$, $m=3$, $\epsilon=10^{-3}$, activity distributed as $F(a)\sim a^{-2.2}$, we plot $I_\infty$ as a function of $\beta/\mu$ and $p$. Red curves represent the critical value $p_c$. Panels D shows the comparison of the stationary state of a SIS model with and without control strategy, $I_\infty^p/I^0_\infty$, as a function of $p$ when $\beta/\mu=0.81$. In green triangles we consider the random strategy, in blue diamonds the targeted strategy, and in orange circles the egocentric strategy. Each plot is made averaging $10^2$ independent simulations started with $1 \%$ of random seeds.}
\label{fig:Fig2}
\end{figure}

\begin{figure}
\includegraphics[width=8cm]{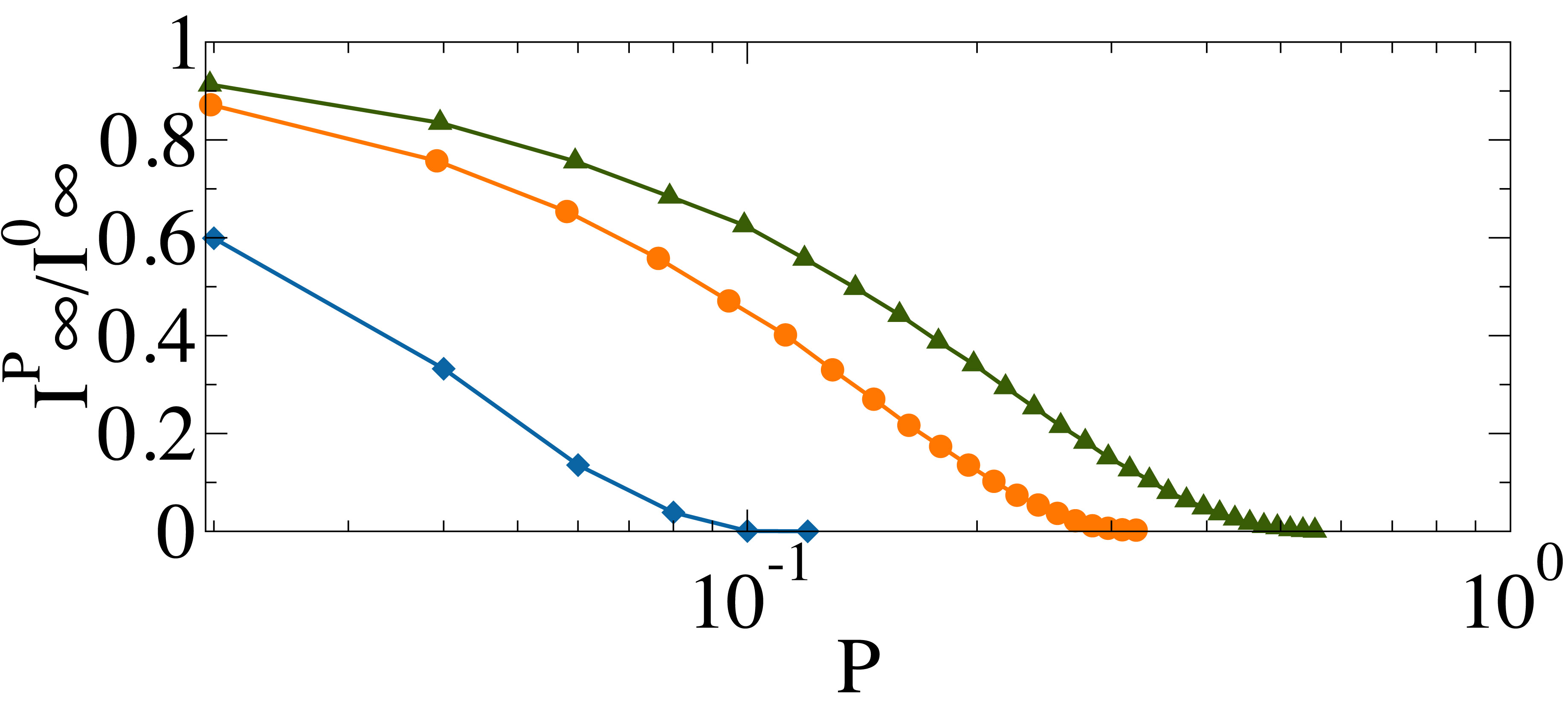}
\caption{ Comparison of the stationary state of an SIS model with and without control strategies. The $I_\infty^p/I^0_\infty$ ratio of final fraction of diseased nodes of random (egocentric, targeted) strategy $I_\infty$, and without intervention $I^0_\infty$ is shown with green triangles (orange circles, blue diamonds) as a function of $p$. Every simulations were initiated with $1\%$ infected seed, executed $10^2$ times for $T=10^4$ step with $\beta/\mu=2.5$. Each step was integrated for $6\times10^2$ seconds and periodic temporal boundary condition was applied.}
\label{fig:Fig3}
\end{figure}

The spreading of infectious diseases, malwares, scientific
ideas, memes are just few example of real world
phenomena that can be modeled as contagion processes
on networks~\cite{barrat08-1,newman10-1,havlin-book, butts09-1}. It has long been acknowledged that
network structures and connectivity patterns are a relevant
factor in determining the properties of spreading
processes. A number of strategies aimed at controlling them have been proposed with the aim of improving on the random removal strategies that originally defined the concept of herd immunity. Those strategies target nodes according  the number of connections (degree), the k-core or the betweenness of nodes, just to mention a few examples~\cite{psatorras02,kitsak10-1}. The efficiency of each strategy is then measured by its effect on the contagion process when a fraction $p$ of nodes is removed from the system. More precisely, the smaller is the fraction $p$ of nodes removed in order to halt the contagion process and the more effective is the strategy.  \\
Although most real world networks show a high level of dynamic activity, the large majority of theoretical results concerning the control of  contagion processes have been obtained by using a complete timescale separation between the contagion process, $\tau_P$, and the change in network's structure, $\tau_G$. In these approaches the dynamical process takes place in either static ($\tau_P \ll \tau_G$) or annealed ($\tau_P \gg \tau_G$) networks. However, when the two time scales are comparable these convenient approximations might introduce uncontrolled biases in the correct characterization of the dynamical properties of the contagion process~\cite{morris93-1,morris07-1,clauset07,alex12-1,Rocha:2010,Isella:2011,Stehle:2011nx,Karsai:2011,Miritello:2011,dynnetkaski2011,albert2011sync,Parshani:2010,Bajardi:2011,panisson11-1,consensus_temporal_nrets_2012,starnini_rw_temp_nets,perra12-1,ribeiro12-2,perra12-2,karsai13-1,liu13-1,lambiotte12-1,toro2007}. \\
Here we investigate the effect of time-varying connectivity pattern of networks on contagion control strategies by considering the specific class of activity driven network models~\cite{perra12-1}. In particular, we consider the susceptible-infected-susceptible (SIS) contagion model~\cite{keeling08-1} and derive analytically its critical immunization threshold in three different control strategies. We also validate the findings obtained in synthetic networks by studying the effect of each strategy in a large-scale mobile telephone call dataset. \\ 
The class of network models we consider is based on the activity rate $a_i$ of each node $i$. This is the probability per unit time to establish interactions with other individuals. The activity rates are assigned according to a given probability distribution $F(a)$.  The generative network process is defined according to the following rules (see Fig.~\ref{fig:Fig1}-A-B): i) At each discrete time step $t$ the network $G_t$ starts with $N$ disconnected vertices; ii) With probability $a_i \Delta t $ each vertex $i$ becomes active and generates $m$ links that are connected to $m$ other randomly selected vertices; iii) At the next time step $t + \Delta t$, all the edges in the network $G_t$ are deleted. All interactions have a constant duration $\tau_G = \Delta t$ defining the time scale describing the evolution of the network. In the following, without loss of generality, we will set $\Delta t=1$. \\
Activity driven networks are random and memoryless. The full dynamics of the network and its ensuing structure is thus completely encoded in the activity distribution $F(a)$. We consider heavy-tailed distributions of activity i.e. $F(a)=Ba^{-\gamma}$ that reproduce the behavior observed in real data for a number of real-world networks~\cite{perra12-1,ribeiro12-2} and more generally found in human behavioral patterns. The following calculations can be easily extended to any functional form of the activity distribution. Although activity driven models  in their simplest formulation do not account for many features such as link persistence, homophily etc.~\cite{morris93-1,morris07-1,clauset07,alex12-1,Rocha:2010,Isella:2011,Stehle:2011nx,Karsai:2011,Miritello:2011,dynnetkaski2011, albert2011sync,Parshani:2010,Bajardi:2011, consensus_temporal_nrets_2012,starnini_rw_temp_nets,panisson11-1}, they allow the analytical formulation of the concurrent network and contagion process dynamics in the form of appropriate mean-field equations, thus allowing the quantitative study of the dynamical process of interest~\cite{perra12-1,perra12-2,ribeiro12-2,karsai13-1,liu13-1}. \\
We study contagion processes in activity-driven models using the basic SIS model~\cite{keeling08-1}. Each node at each time $t$ can be in the susceptible, $S^t$, or infectious, $I^t$, state. The basic SIS rules thus define a reaction scheme of the type $S+I\to 2I$ with reaction rate $\beta$ and $I\to S$ with reaction rate $\mu$, which represent the contagion and recovery processes, respectively.  A central concept of contagion phenomena is the epidemic threshold. It defines the conditions necessary for the macroscopic spreading of the disease. In networks, the threshold depends on the moments of the degree distribution $P(k)$ that specify the probability that any node is connected to $k$ distinct nodes.  In uncorrelated annealed networks the threshold condition reads  $\frac{\beta}{\mu}\ge \frac{\av{k}^2}{\av{k^2}}$, where $\av{k}$ and $\av{k^2}$ are the first and second moment of the degree distribution, respectively~\cite{alex12-1}. In this expressios  $\beta=\lambda \av{k}$ is the per capita spreading rate that takes into account the rate of contacts $\av{k}$. While the expression might be different in the case of static networks~\cite{wang03,castellano10-1,durrett10-1} the topological properties of the underlying network have critical effects on the threshold.
In time-varying networks the analytical study of  contagion processes is hindered by the difficulties in dealing with the concurrent time scales of the contagion and network evolution processes.~\cite{prakash10-1,starnini13-1,lee12-1,takaguchi12-1,tang11-1,Masuda13-1}. In the case of activity driven networks however it is possible to derive the mean-field level dynamical equations describing the contagion process by defining the activity block variable $I^{t}_{a}$ and $S^{t}_{a}$ that represent the number of infected and susceptible individuals, respectively, in the class of activity $a$ at time $t$. From those quantity it is possible to derive the mean-field evolution of the number of infected individuals of class $a$  at time $t + 1$ as
\begin{eqnarray}
\label{eq_1}
I^{t+1}_{a} &=& I^{t}_{a} -\mu I^{t}_{a}+ m (N_{a}-I_{a}^{t}-R_a^t)a \int d a'  \frac{I_{a'}^{t}}{N} + \\ \nonumber
&+&\lambda  m(N_{a}-I_{a}^{t}-R_a^t)\int d a'  \frac{I_{a'}^{t}a'}{N},
\end{eqnarray}
where $N_a$ is the total number of individuals with activity rate $a$  (constant over time) and $R_a^t$ are the nodes in the class $a$ at the time $t$ that have been immunized/removed from the network. In Eq.~\ref{eq_1} the first term considers the number of infected in class $a$ at time $t$. The second term describes the number of nodes that recover going back in the class $S_a$.  The third term instead  the number of infected individuals generated when nodes in the class $S^t_a=N_a-I^t_a-R_a^t$ are active and connect with infected nodes in the other activity classes. Finally, the last term considers the number of infected generated when nodes in the class $S^t_a$ are linked by active infected nodes in other activity classes. In the case of a random immunization nodes are selected independently of their features. In absence of any controlling strategy the $R_a^t=0$. In this case, summing on all the classes and ignoring  the second order terms in $I/N$ we can write $I^{t+ 1}=I^{t}- t\mu  I^{t} + \lambda m \langle a \rangle I^{t}+ \lambda m \theta^{t}$, where $\theta^{t}=\int da' I^{t}_{a'}a'$. Multiplying both sides of by $a$ and integrating we obtain $\theta^{t+1}=\theta^{t}-\mu  \theta^{t}+ \lambda  m \langle a^2 \rangle I^{t}+ \lambda  m \langle a \rangle\theta^{t}$.
The system equations for $I^{t+ 1}$ and $\theta^{t+1}$ provides an epidemic outbreak only if the eigenvalues of the corresponding matrix is larger than 1 so that the epidemic $\xi$ threshold reads as $\frac{\beta}{\mu} \ge \xi^{SIS}\equiv \frac{2\langle a \rangle }{\langle a \rangle +\sqrt{\langle a^2 \rangle}}$, 
where $\langle a \rangle$ and $\langle a^2 \rangle$ are the first and second moment of the activity distribution (see Ref.~\cite{perra12-1} and the SI for the calculation details). Remarkably, the threshold does not depend on the time-aggregated network representation, and it is just a function of the interaction rate of the nodes. \\
By using this framework we can study different immunization strategies. Let us consider first the random strategy (RS) in which a fraction $p$ of nodes is immunized with an uniform probability (see Fig.~\ref{fig:Fig1}-D). In this case the system of equations describing the dynamic of the system can be obtained by setting $R_a=p N_a$. The epidemic threshold condition changes as the eigenvalue expression of the system of dynamical equations is a function of the fraction $p$, finally yielding
\be
\label{thre_RI}
\frac{\beta}{\mu} \ge \xi^{RS} \equiv \frac{1}{1-p}\frac{2\av{a}}{\langle a \rangle +\sqrt{ \langle a^2 \rangle}} =\frac{\xi^{SIS}}{1-p}.
\ee
As expected, when a fraction $p$ of nodes is randomly immunized/removed the epidemic threshold can be written as the threshold with no intervention, $\xi^{SIS}$, rescaled by the number of nodes still available to the spreading. Indeed immunizing/removing random nodes is equivalent to rescale the per capita spreading rate by the fraction of available nodes $\beta \rightarrow \beta(1-p)$ (see also the SI material).  Another important quantity is the critical value of immunized/removed nodes, $p_c$, necessary to halt the contagion process. This quantity is a function of the network's structure, and the specific features of the contagion process.  The explicit value of $p_c$ can be obtained by inverting Eq.~\ref{thre_RI}, and in Fig.~\ref{fig:Fig2}-A we plot $p_c$ as a function of $\beta/\mu$ keeping fixed the statistical properties underlying network. We then simulated the process for each pair of values and plot the average asymptotic density of infected nodes, $I^p_{\infty}$,  in $10^2$ independent realizations. The phase space of the diffusion process is divided in two different regions separated by the red solid line that represents $p_c$ as derived by Eq.~\ref{thre_RI}.  The region below the curve is the region in which the spreading process will take over, $p<p_c$. The region above the curve is instead characterized by all the value of $p$ larger/equal than $p_c$. In this region the fraction of removed/immunized nodes is enough to stop completely the diffusion process. To further assess the efficiency of the immunization strategy in Fig.~\ref{fig:Fig2}-D (green triangles) we plot, as a function of the density of removed/immunized nodes $p$, the ratio $I_{\infty}^p/I^0_{\infty}$ where $I^0_{\infty}$ is the asymptotic density of infected nodes when no-intervention is implemented. As clear from the figure, the random strategy allows to decrease the fraction of infected nodes just for large values of $p$ confirming its inefficiency also in time-varying networks. \\
In networks with heavy-tailed degree distribution strategies  targeting the removal of nodes with high degree centrality performs more efficiently than random strategies~\cite{barrat08-1,havlin-book}. In activity driven networks analogous strategies shall target high activity nodes. For this reason, we rank nodes in decreasing order of activity and the first $pN$ nodes are immunized/removed,  see Fig.~\ref{fig:Fig2}-E. This method is equivalent to fix a value $a_c$ so that any node with activity $a\ge a_c$ is immune to the contagion process~\footnote{The value of $p$ and $a_c$ are linked by the relation $p=\int_{a_c}^{1}F(a)da$}. Also for this strategy it is possible to derive the analytic expression for the epidemic threshold (see SI): 
\be
\label{thre_targ}
\frac{\beta}{\mu}\ge \xi^{TS}\equiv\frac{2 \langle a \rangle}{\langle a \rangle^{c}+\sqrt{(1-p) \langle a^2 \rangle^{c} }},
\ee
where $ \xi^{TS}$ indicates the threshold in the case of the targeted strategy control strategy. In the above expression we define $\av{a^n}^c=\int_{\epsilon}^{a_c}a^{n}F(a)da $ the moments of the activity distribution discounting the removed/immunized nodes. Eq.~(\ref{thre_targ}) is not a simple rescaling of the original threshold expression and implies a drastic change in the behavior of the contagion process.  As shown in details in the SI the analytical predictions (Eq.~\ref{thre_targ}) well reproduce the behavior observed in numerical simulations. In order to define the critical value of $p$ necessary to completely stop the spreading we have to invert Eq.~\ref{thre_targ}.  The moments of the distribution of the remaining nodes are function of $p$ through $a_c$ and it is not possible to derive explicitly $p_c$. However, it can be easily evaluated numerically by solving the equation $\xi^{TS}-\beta/\mu=0$ for different values of $\beta/\mu$. In Fig.~\ref{fig:Fig2}-B we show $p_c$ (red line) as a function of $\beta/\mu$. The efficiency of the targeted strategy is clear. Immunizing/removing a very small fraction of the most active nodes results is enough to stop the contagion process. This is confirmed in Fig.~\ref{fig:Fig2}-D (blue diamonds) where we plot the ratio $I_\infty^p/I^0_\infty$.  The extreme efficiency of this strategy is due to crucial role of high activity nodes in the spreading process in driving the diffusion. By immunizing just 1\% of the highest activity nodes the epidemic  processes dies out.\\
Unfortunately the network-wide knowledge required to implement targeted control strategies is generally not available~\cite{cohen03-1}. In the case of evolving networks this issue is even more pronounced as node's characterization is depending on how long it is possible to observe the network dynamics.  Among the different possible approaches aimed at overcoming this issue, we consider a strategy based on sampling the egocentric network of fraction $w$ of nodes randomly selected. These nodes act as ``probes".  During an observation time $T$ each probe might have interactions with other nodes. Among them one is randomly selected and immunized/removed (see Fig.~\ref{fig:Fig1}-F). For the sake of comparison with the previous control strategies, we define the fraction of actual immunized nodes as $p$. In general $w\neq p$~\footnote{In order to guarantee that a fraction $w$ of nodes is immunized/removed the systems need to be observed for more than one time step. We define $T^*$ as the average time needed for all the probes to have  at least one interaction with other nodes. For any observation time $T< T^*$ the fraction of immunized/removed nodes will be in general $p\le w$.}. In this scheme the probability of immunization/removal for one node with activity $a$ after a time step is: 
\be
\label{p_a}
P_a=a  w\int d a'  \frac{m N_{a'}}{N} + w\int d a' a'\frac{m N_{a'}}{N}\frac{1}{m}.
\ee
The first term on the right side considers the probability that a node of class $a$ is active and reaches one of the probes; the second term instead, the probability that one node of class $a$ gets a connection from one active probe. Solving the integrals in Eq.~(\ref{p_a}) we can write $P_a= w \left ( am+\langle a \rangle \right)$. The probability of immunization of one node in the activity class $a$ after $t$ time steps is $P_a^t=1-(1-P_a)^t$ and therefore summing over all the activity classes we can estimate the total number of immunized individuals as $R^T = \sum_{a}N_aP_a^T=\sum_{a}N_a \left [ 1-(1-P_a)^T \right ]$. The equation for $P_a$ does not consider the depletion of nodes in each class due to the immunization process. The formulation is then a good approximation for small $w$ and $T$, when the probability that a probe is selected more than one time is very small. By replacing the expression for the removed/immunized individuals in the basic SIS equations (see SI) yields the following epidemic threshold for the egocentric sampling strategy (ESS): 
\be
\label{thre_ES}
\frac{\beta}{\mu} \ge \xi^{ESS} \equiv \frac{2\av{a}}{\Psi_1^T +\sqrt{ \Psi_0^T \Psi_2^T}},
\ee
where we define  $\Psi_{n}^T=\int da \,\ a^{n}(1-P_a)^TF(a)$. This last integral is a function of the observing time window $T$, the probability of immunization/removal of each class and the distribution of activity.  We evaluate  each $\Psi$ term through numerical integration. As shown in the SI the analytical predictions (Eq.~\ref{thre_ES}) nicely reproduce the behavior observed in numerical simulations.
As done for the other two measures we define the critical value of $p$ by solving numerically the equation $\xi^{ESS}-\beta/\mu=0$ for different values of $\beta/\mu$. In Fig.~\ref{fig:Fig2}-C we show $p_c$ (red solid line) as a function of $\beta/\mu$. From the plot is clear how this strategy is much more efficient than the random one, although not as performant as the targeted one (see also Fig.~\ref{fig:Fig2}-D). The efficiency of this strategy is due to the ability to reach out most active nodes by a local exploration done observing the systems for few time steps.\\
Real world time-varying networks add a number of complications to the simplified picture offered by activity driven networks. Indeed they exhibit correlations among nodes, persistency of links, burstiness of the activity pattern, just to cite a few~\cite{morris93-1,morris07-1,clauset07,alex12-1,Rocha:2010,Isella:2011,Stehle:2011nx,Karsai:2011,Miritello:2011,dynnetkaski2011, albert2011sync,Parshani:2010,Bajardi:2011, consensus_temporal_nrets_2012,starnini_rw_temp_nets,panisson11-1}. It is therefore extremely important to validate, at least in its basic phenomenology, the above mean-field framework in real world datasets. We consider a mobile phone call data network constituted of $93,190$ connected phone users of a single city involved in almost five millions calls over $120$ days. The node activity is time-stamped and the full network time-varying pattern is available. We simulate the  SIS spreading process imposing $1\%$ of randomly chosen infectious nodes and implementing the three immunization strategies previously described (see SI for details on the numerical simulations). We report in Fig.~\ref{fig:Fig3} the behavior of $I_\infty^p/I^0_\infty$, providing a measure of the effectiveness of each strategy. The agreement between what observed in real time-varying networks characterized by non-Markovian dynamics, and the analytical results for activity driven networks is remarkable. \\
In summary, we have present an analytical framework that offers a general picture of the behavior of control strategies for contagion processes  in both synthetic and real world time-varying networks. The theoretical approach presented in this paper can potentially include more complicated time-varying patterns offering a general analytical  and interpretative framework for the understanding of  how we can control and mitigate (as well as enhance) contagion phenomena by acting on the basic parameters of the system. Such results find applications in a broad range of real world phenomena ranging from the diffusion of emerging diseases to information and communication processes.\\
This work has been partially funded by the NSF CCF-1101743 and CMMI-1125095 awards to AV. We thank A.-L. Barab\'asi for the validation dataset.


\section{Supporting Information}

\subsection{ Activity-driven networks}

Activity driven models are based on the activity rate $a_i$ of each node $i$. This  is the probability per unit time to establish interactions with other individuals. The activity rates are assigned according to a given probability distribution $F(a)$.
The activity driven model uses the
activity distribution to drive the formation of a dynamic network. 
In particular $N$ disconnected nodes are initially considered. In this setting the generative network process is defined
according to the following rules:
\begin{itemize}
\item At each discrete time step $t$ the network $G_t$ starts with $N$
  disconnected vertices;
\item With probability $a_i \Delta t$ each vertex $i$ becomes active
  and generates $m$ links that are connected to $m$ other randomly
  selected vertices. Non-active nodes can still receive connections
  from other active vertices;
\item At the next time step $t + \Delta t$, all the edges in the
  network $G_t$ are deleted. 
 \end{itemize}
 
From this definition it follows that all interactions have a constant duration $\tau_G = \Delta t$. This defines the time scale describing the evolution of the network structure. In the following, without loss of generality, we will set $\Delta t=1$. Activity driven model is random and memoryless. Nodes do
not have memory of the previous time steps. The full dynamics of the
network and its ensuing structure is thus completely encoded in the
activity distribution $F(a)$.
If not specify otherwise we consider heavy-tailed distributions of activity i.e. $F(a)=Ba^{-\gamma}$ that reproduces behavior similar to what is observed in real data~\cite{perra12-1,ribeiro12-1} for a number of real-world networks. 

The model allows for a simple analytical treatment. Let us define the
integrated network $G_T=\bigcup_{t=0}^{t=T} {G}_t$ as the union of all
the networks obtained at each previous time step. The instantaneous
network generated at each time $t$ will be composed of a set of
slightly interconnected nodes corresponding to the nodes that were
active at that particular time, plus those who received connections
from active nodes. The average degree of a node $i$ characterized by an activity rate $a_i$ will be:
\be
\av{k_i}_t= m a_i + m\av{a}.
\ee
The first contribution is due to the $m$ links generated by the node when active. The second contribution is due to the links that reach node $i$ coming from other active nodes. At each time step the average degree in the network will be
\be
\av{k}=2m\av{a}.
\ee 

Taking a snapshot of the instantaneous network we will see a set of
stars, the vertices that were active at that time step, with degree
larger than or equal to $m$, plus some vertices with low degree. The
corresponding integrated network, on the other hand, will generally
not be sparse. It is the union of all the instantaneous networks at
previous times. In particular it is possible to show that the degree distribution
$P_T(k)$ of the integrated network at time $T$ takes the form:
\begin{equation}
P_T(k)\sim \frac{1}{T m \eta}F\left[ \frac{k}{T m \eta}\right], 
\end{equation}
in the limit of small $k/N$ and $k/T$~\cite{perra12-1}. This is an important feature of the model. 
The integrated degree distribution has the same function form of the activity. Fixing an opportune distribution for $F(a)$, the model could reproduce the connectivity distribution observed in the data and consider explicitly the dynamical evolution of its structure. \\

\subsection{Threshold Behavior of Contagion Models in Time-Varying Networks}

Here we study the $SIS$ and $SIR$ model~\cite{kermac27-1,keeling08-1,barrat08-1} on time-varying networks. The system is divided in two different classes for $SIS$ processess and three for $SIR$ processess. The class $S$ represent nodes that are susceptible to the propagation. The class $I$ represent nodes that are infected/informed and propagate the disease/rumor. The class $R$ of nodes represent nodes that are recovered and cannot be infected again. Node in the class $S$ will move to the class $I$  with infection rate $\lambda$ if it is connected with a node in the class $I$.  Nodes stay in the  class $I$ an average time $\mu^{-1}$  which is defined as the inverse of the recovery rate $\mu$. In the case of the $SIS$, nodes go back to the $S$ class returning to be susceptible to the spreading. In the $SIR$ instead they enter to the class $R$, become recovered and cannot be infected again. Indeed in these cases the condition $\lambda>0$ is necessary but not sufficient to guarantee the spreading. There is an interplay between the infection rate and the recovery rate~\cite{kermac27-1,keeling08-1,barrat08-1}. These processes are characterized by a threshold that separate the phase space in two regions: one characterized by a finite fraction of the nodes reached by the spreading and the other instead where the process does not take over and dies.

\subsection{$SIS$ processess}
The process is characterized by two transitions between the two different states:
\begin{eqnarray}
S+I &\xrightarrow{\lambda}& 2I, \\
I &\xrightarrow{\mu}& S.
\end{eqnarray}
While the first transition requires an interaction between a node in the class $S$ and one in the class $I$, the second is spontaneous. The stationary state of the process is characterized by two different behaviors. Above threshold an endemic is reached where the density of nodes in the class $I$, $I_\infty$,  is fixed by the balance of the spreading and recovery rate. Below threshold the number of nodes in the class $I$ is zero. The spreading is not able to sustain itself.

Let us consider $N$ nodes characterized by activity $a$ extracted from a general distribution $F(a)$. In the activity driven framework nodes are grouped by their activity and nodes in the same class are statistically equivalent. At the mean-field level, the spreading process can be described by the number of infected individuals in the class of activity $a$ at time $t$, namely 
$I^{t}_{a}$.  The number of infected individuals of class $a$  at time $t + 1$ given by:
\begin{eqnarray}
\label{pp1}
I^{t+1}_{a} &=& I^{t}_{a} -\mu I^{t}_{a}   +\lambda  m (N_{a}-I_{a}^{t})a   \int d a' \frac{I_{a'}^{t}}{N}  \\ \nonumber
&+&\lambda  m(N_{a}-I_{a}^{t})\int d a' \frac{I_{a'}^{t}a' }{N},
\end{eqnarray}
where $N_a$ is the total number of individuals with activity rate $a$  (constant over time).  Each term in the Eq.~(\ref{pp1}) has a clear physical interpretation. The number of infected in the class $a$ at time $t+ 1$ is given by: the number of infected in this class at time $t$ (first term), minus the number of nodes that recover going back to the class $S_a$ (second term), plus the number of infected individuals generated when nodes in the class $S^t_a=N_a-I^t_a$ are active and connect with infected nodes in the other activity classes (third term), plus the number of infected generated when nodes in the class $S^t_a$ are linked by active infected nodes in other activity classes. \\
Summing on all
the classes and ignoring  the second order terms we can write:
\be
\label{pp}
\int da I_{a}^{t+ 1}= I^{t+ 1}=I^{t}-\mu  I^{t} +\lambda m \langle a \rangle I^{t}+  \lambda m \theta^{t},
\ee
where $\theta^{t}=\int da' I^{t}_{a'}a'$. Multiplying both sides of   Eq.~(\ref{pp1}) by $a$ and integrating we obtain:
\be
\theta^{t+ 1}=\theta^{t}- \mu  \theta^{t}+ \lambda  m \langle a^2 \rangle I^{t}+ \lambda  m \langle a \rangle\theta^{t}.
\label{othereq}
\ee
In the continuous time limit, we can write Eqs.~(\ref{pp1}) and~(\ref{othereq}) in a differential form:
\begin{eqnarray}
 \partial_{t} I &=&-\mu I + \lambda m  \langle a \rangle I+ \lambda m \theta,\\
\partial_{t} \theta &=& -\mu\theta +\lambda m \langle a^2\rangle I+\lambda m \langle a \rangle\theta.
\end{eqnarray}
The Jacobian matrix of this set of linear differential equations takes the form
\[ J= \left( \begin{array}{cc}
-\mu +\lambda m \langle a \rangle & \lambda  \\
\lambda m \langle a^2 \rangle & -\mu +\lambda m \langle a \rangle  \\ \end{array} \right), 
\] 
and has eigenvalues
\be
\Lambda_{(1,2)} = \langle a \rangle\lambda m  - \mu \pm \lambda m  \sqrt{\langle a^2\rangle}.
\ee
The threshold  is obtained requiring the largest eigenvalues to be larger the $0$:
\be
\label{t_m1}
\frac{\lambda}{\mu} \ge \frac{1}{m}\frac{1}{\langle a \rangle+\sqrt{\langle a^2 \rangle}}.
\ee
Considering the per capita spreading rate $\beta=\lambda \av{k}$ we can write threshold for the $SIS$ process, $\xi^{SIS}$, as:
\be
\label{thre_SIS}
\frac{\beta}{\mu} \ge \xi^{SIS}\equiv \frac{2\langle a \rangle }{\langle a \rangle +\sqrt{\langle a^2 \rangle}}.
\ee

\subsubsection{$SIR$ processess}

In the $SIR$ process nodes in the class $I$, after recovering, enter the class $R$. These nodes cannot be infected again. The dynamics of the process is thus defined by two transitions:
\begin{eqnarray}
S+I &\xrightarrow{\lambda}& 2I, \\
I &\xrightarrow{\mu}& R.
\end{eqnarray}

For quenched and annealed networks the thresholds of SIS and $SIR$ are slightly different~\cite{barrat08-1}. This is due to the fact that in the $SIR$ model a node of degree $k$ can infect at most  $k-1$ nodes. At least one of its neighbors need to be infected already. It will recover and will not be infected again. In the $SIS$ model instead all $k$ neighbors can be infected. Multiple infections are part of the dynamics. This constraint is relevant when the topology is fixed (quenched) or fixed just on average (annealed). In activity driven networks the connections are memoryless and they change at each time step. This constraint is then less relevant especially in the first phases of the spreading when the infected is small. The thresholds of the two processes in this case are the same. The number of infected individuals in the activity class $a$ at the time step $t+1$ for a $SIR$ process is:
\begin{eqnarray}
I^{t+1}_{a} &=& I^{t}_{a} -\mu   I^{t}_{a}+\\ \nonumber
&+&\lambda  m (N_{a}-I_{a}^{t}-R_a^t)a  \int d a'  \frac{I_{a'}^{t}}{N} + \\ \nonumber
&+&\lambda  m(N_{a}-I_{a}^{t}-R_a^t)\int d a'  \frac{I_{a'}^{t}a' }{N},
\end{eqnarray}
where $R_a^t$ are the nodes in the class $a$ at time $t$ that acquired immunity. In the early stages of the spreading all the terms $(I_a^t)^2$ and $I_a^t R_a^t$ are second order terms. We can write:
\be
\label{pp2}
\int da I_{a}^{t+1}= I^{t+1}=I^{t}- \mu  I^{t} + \lambda m \langle a \rangle I^{t}+   \lambda m \theta^{t},
\ee
that is the same equation we wrote for the $SIS$ process. This confirms the intuition that the two thresholds are the same.

\begin{figure}[h]
\begin{center}
\includegraphics*[width=0.5\textwidth] {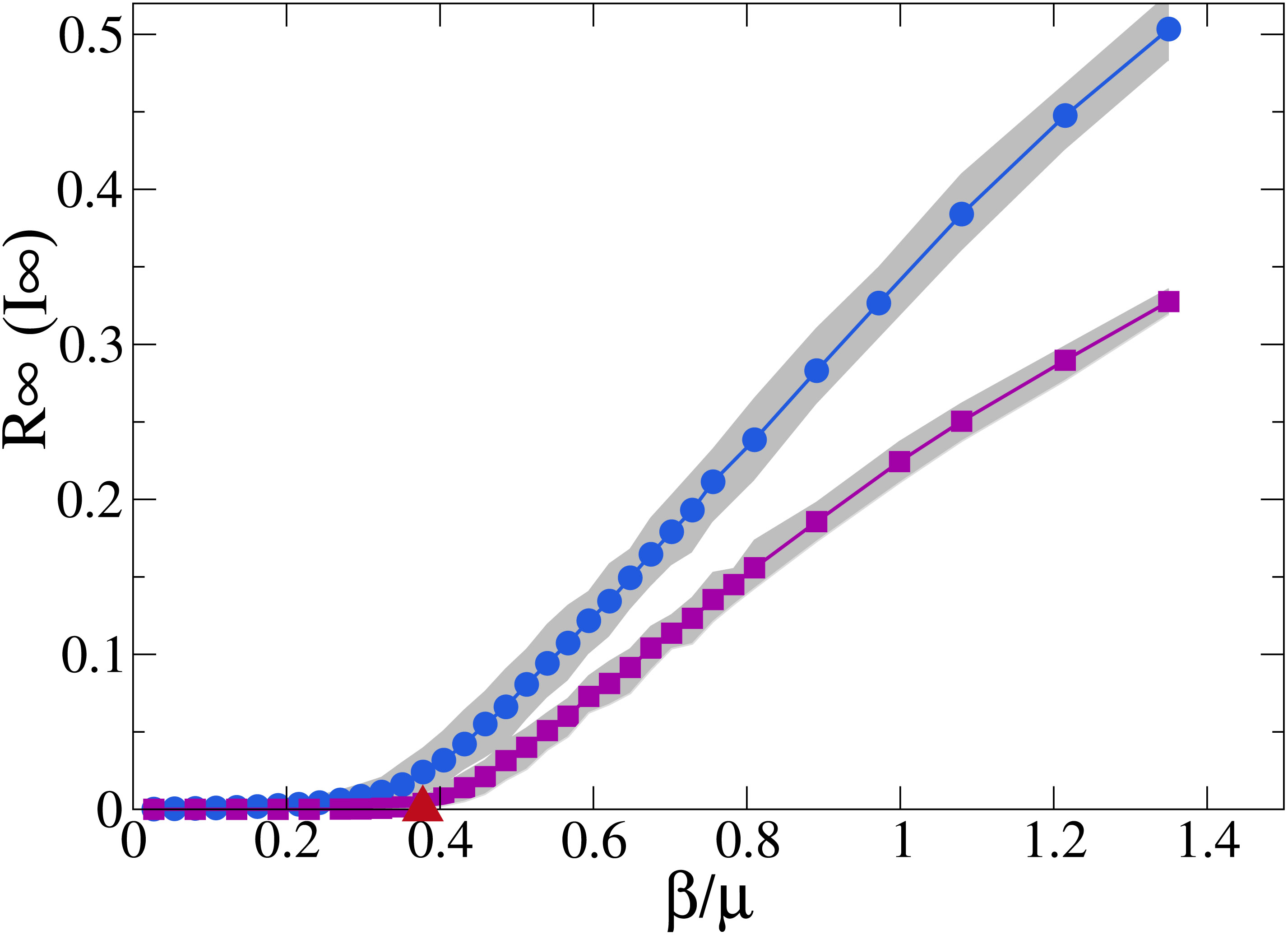}
\caption{ The plot shows the asymptotic density of a $SIR$ ($R_\infty$) and a $SIS$ ($I_\infty$), as a function of $\beta/\mu$. We consider $N=10^5$ nodes, $m=3$, $\epsilon=10^{-3}$, and a power law distribution of activity $F(a)\sim a^{-\gamma}$ with $\gamma=2.2$. In blue circles we plot the median of $R_\infty$ and purple squares represent the median of $I_\infty$. The red triangle represents the analytical prediction of the threshold according to Eq.~\ref{thre_SIS}. In grey are plotted the $95\%$ reference range. The plot is made averaging $10^2$ independent simulations and each one of them starts with $1 \%$ of random seeds. }
\label{sir}
\end{center}
\end{figure}

In Figure~\ref{sir} we plot the asymptotic value of nodes in the class $R$ for $SIR$ and in the class $I$ for $SIS$ as a function of $\beta/\mu$. The red triangle represent the analytical prediction for the $SIS$ process according to Eq.~\ref{thre_SIS}. The numerical simulations show that in activity driven networks the thresholds for $SIS$ and $SIR$ models are the same. In particular we can argue that control strategies for the spreading process will affect in the same way the thresholds of the SIS and SIR model.
For this reason in the following we will provide general analytical results valid for both models and limit our numerical investigations to the SIS model.

\subsection{Control Strategies in Time-Varying Networks}

We consider the following three containment strategies to select a fraction $p$ of nodes that will be immunized/removed and will not be able to participate in the spreading process. 

\subsubsection{Random Strategy }

In this strategy a fraction $p$ of random nodes is immunize/removed. These nodes cannot participate to the spreading process. Let us consider $N$ nodes and the distribution of activity $a$ of the nodes is given by a general 
distribution $F(a)$ as before. Let us define the number of immunized/removed nodes in each activity class as $R_{a}$.
The number of infected individuals of class $a$  at time $t+1$ given by:
\begin{eqnarray}
\label{im_r1}
I^{t+1}_{a}&=&  I^{t}_{a}-\mu   I^{t}_{a} +\\ \nonumber
 &+&\lambda (N_{a}-I_{a}^{t}-R_{a})a   \int d a' \frac{I_{a'}^{t} m }{N}+ \\ \nonumber
 &+&\lambda (N_{a}-I_{a}^{t}-R_{a})\int d a' \frac{I_{a'}^{t}a'  m }{N}.
\end{eqnarray}
The form of this equation is the same as before with the difference that  the number of susceptible is each activity class $a$ is $S^t_a=N_a-I^t_a-R_a$. In this strategy each node has the same probability $p$ to be selected, so we can write that $R_a=p N_a$. Using this in previous equation and following the same lines described before we can write the threshold for the random strategy $\xi^{RS}$:
\be
\label{thre_RI}
\frac{\beta}{\mu} \ge \xi^{RS} \equiv \frac{1}{1-p}\frac{2\av{a}}{\langle a \rangle +\sqrt{ \langle a^2 \rangle}} =\frac{\xi^{SIS}}{1-p}.
\ee

\begin{figure}[h]
\begin{center}
\includegraphics*  [width=0.5\textwidth] {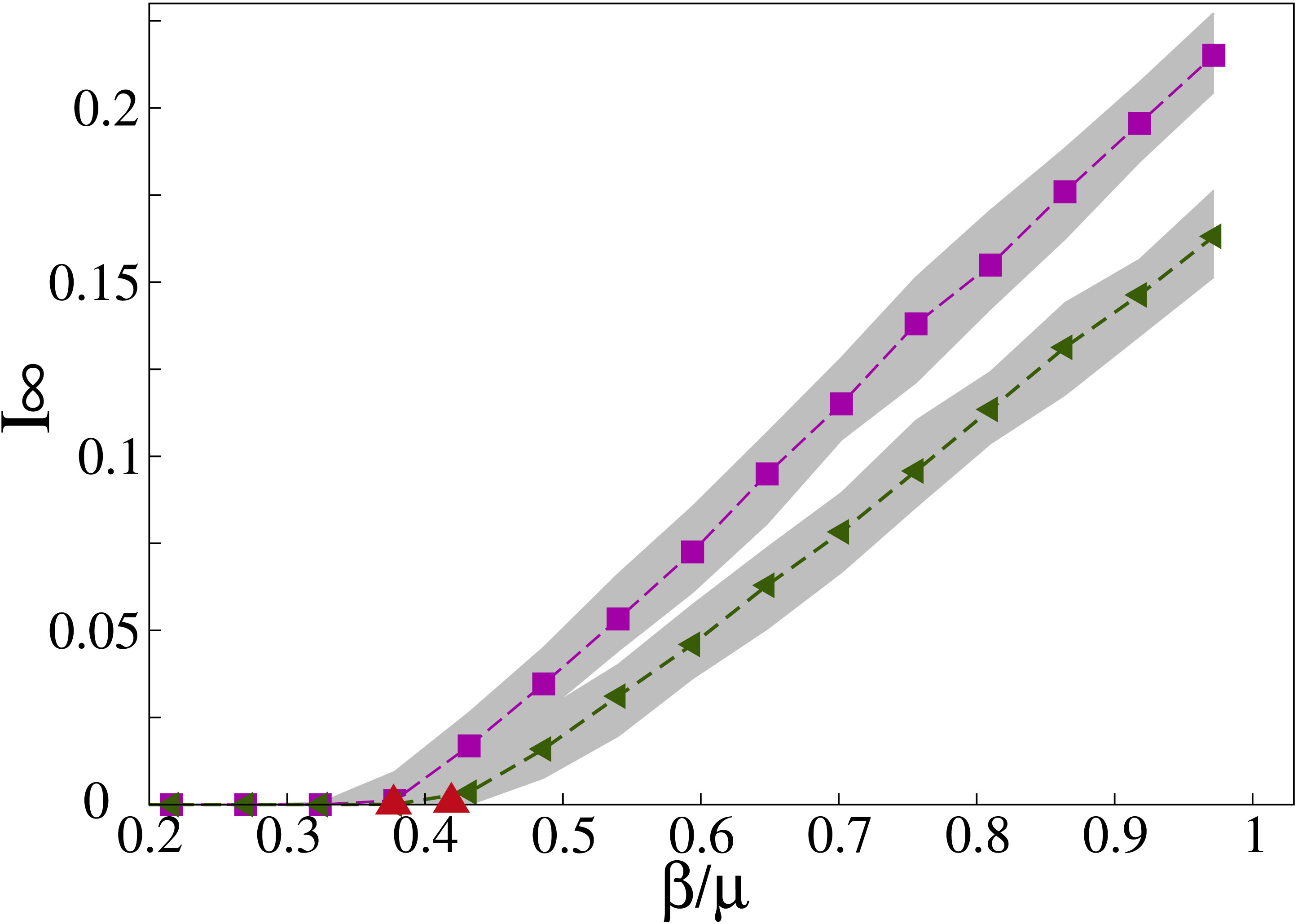}
\caption{ The plot shows the asymptotic density of a SIS model, $I^p_\infty$, as a function of $\beta/\mu$. We consider $N=10^5$ nodes, $m=3$, $\epsilon=10^{-3}$, and a power law distribution of activity $F(a)\sim a^{-\gamma}$ with $\gamma=2.2$. In purple squares we plot the median of $I^p_\infty$ when no intervention is implemented. In green triangles we plot the median of $I^p_\infty$ when a random strategy with $p=0.1$ of intervention is implemented. The red up triangles represent the analytical prediction of the thresholds according to Eq.~\ref{thre_SIS} and Eq.~\ref{thre_RI}. In grey are plotted the $95\%$ reference range. The plot is made averaging $10^2$ independent simulations and each one of them starts with $1 \%$ of random seeds.  }
\label{sis_random}
\end{center}
\end{figure}

In Figure~\ref{sis_random} we plot the asymptotic value of the density of nodes in the class $I$ as a function of $\beta/\mu$. We plot the $I^p_\infty$ when the random strategy is implemented (green triangles) compared to the baseline when no intervention is applied (purple squares). The red up triangles represent the analytical prediction of the thresholds according to Eq.~\ref{thre_SIS} and Eq.~\ref{thre_RI}. The plot confirms that the analytical predictions for both thresholds are nicely reproduced by the simulations. 

An important quantity to evaluate is the critical value of immunized/removed nodes, $p_c$, necessary to completely stop the process. The critical value $p_c$ is a function of networks' structure and of the characteristics of the spreading process. Inverting Eq.~\ref{thre_RI} we can write:
\be
\label{pc_rand}
p_c=1-\frac{2\av{a}}{\av{a}+\sqrt{\av{a^2}}}\frac{\mu}{\beta}.
\ee

\subsubsection{Targeted Strategy }

In this strategy nodes are ranked in decreasing order of activity. The first $pN$ are immunized/removed. This method is equivalent to fix a value $a_c$ so that any node with activity $a\ge a_c$ is protected. The value of $p$ and $a_c$ are easily related:
\be
p=\int_{a_c}^{1}F(a)da.
\ee
Considering a power law distribution of activity $F(a)=Aa^{-\gamma}$ we can solve the integral and write:
\be
p=\frac{1}{1-\epsilon^{1-\gamma}}(1-a_c^{1-\gamma}).
\ee
All the nodes in activity class larger or equal to the cut $a_c$ do not take part to the spreading. The process is sustained by all the other activity classes. The equation that describes the variation in the number of infected nodes for a given class $a<a_c$ is then:
\begin{eqnarray}
I^{t+1}_{a} &=& I^{t}_{a} -\mu   I^{t}_{a} + \\ \nonumber 
&+&\lambda  (N_{a}-I^t_a)a   \int_{\epsilon}^{a_c} d a' \frac{I_{a'}^{t} m }{N}+\\ \nonumber
&+&\lambda   (N_{a}-I^t_a)\int_{\epsilon}^{a_c} d a' \frac{I_{a'}^{t}a'  m }{N}.
\end{eqnarray}
The form of the equation is the same written as no intervention strategies are implemented. The differences are in the integrals that now run in the region of allowed activities. Following the same lines as before we can write the threshold for this process with targeted immunization/removal of nodes, $\xi^{TS}$:
\be
\label{thre_targ}
\frac{\beta}{\mu}\ge \xi^{TS}\equiv\frac{2 \langle a \rangle}{\langle a \rangle^{c}+\sqrt{(1-p) \langle a^2 \rangle^{c} }},
\ee
where we define:
\be
\label{mom}
\av{a^n}^c=\int_{\epsilon}^{a_c}a^{n}F(a)da,
\ee
as the moments of the activity distribution in the region of not suppressed nodes. The functional form of the threshold is similar to the baseline with the difference that it is now driven by the activity of the nodes that are sustaining the spreading. 

\begin{figure}[h]
\begin{center}
\includegraphics*  [width=0.5\textwidth] {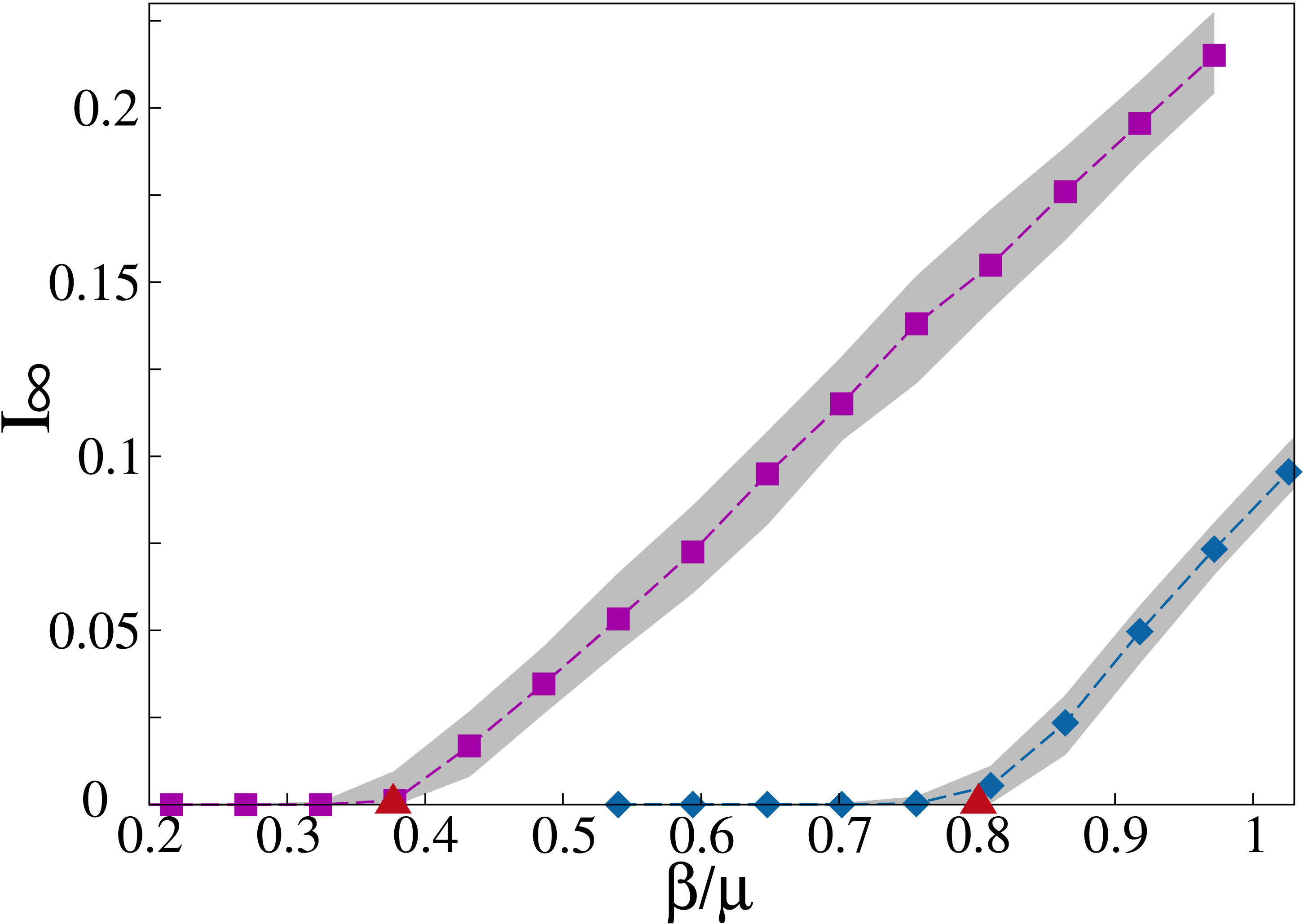}
\caption{ The plot shows the asymptotic density of a SIS model, $I_\infty$, as a function of $\beta/\mu$. We consider $N=10^5$ nodes, $m=3$, $\epsilon=10^{-3}$, and a power law distribution of activity $F(a)\sim a^{-\gamma}$ with $\gamma=2.2$.  In purple squares we plot the median of $I_\infty$ when no intervention is implemented. In blue diamonds we plot the median of $I_\infty$ when a targeted strategy of intervention is implemented. We fixed $p=3\times 10^{-3}$ (or $a_c=0.1$). The red up triangles represent the analytical prediction of the thresholds according to Eq.~\ref{thre_SIS} and Eq.~\ref{thre_targ}. In grey are plotted the $95\%$ reference range. The plot is made averaging $10^2$ independent simulations and each one of them starts with $1 \%$ of random seeds.}
\label{sis_target}
\end{center}
\end{figure}

In order to have a clear understanding of these differences, in Figure~\ref{sis_target} we plot the asymptotic density of nodes in the class $I$, $I_\infty$, for different value of $\beta/\mu$ when no interventions are applied (purple squares) and when the targeted strategy is implemented (blue diamonds). The threshold $\xi^{TS}$ is more than two times larger than the baseline just considering a small fraction of protected nodes $p = 3 \times 10^{-3}$ (corresponding to $a_c=0.1$) as predicted from the analytical approach. \\

\subsubsection{Egocentric Sampling Strategy }

In this method a fraction of $\omega$ nodes is selected at random and each one of them is asked to point one of its neighbors. The neighbors are then immunized/removed. The initial pool of random selected nodes act as probes of the system. In activity driven networks, at each time step just few nodes are active and have connections. In order to guarantee that a fraction $p$ of nodes is immunized/removed, the system needs to be observed for more than one time step. We define $T^*$ as the average time needed for all the probes to have at least one connection and being able to indicate one of their neighbors for protection. Observing the network for $T< T^*$ the fraction of immunized/removed nodes will be in general $p\le \omega$. 

In the activity driven framework the probability of immunization/removal for one node with activity $a$ after a single time step is: 
\be
\label{p_a}
P_a=a   \omega \int d a'  \frac{m N_{a'}}{N} + \omega \int d a' a' \frac{m N_{a'}}{N}\frac{1}{m}.
\ee
The first term on the right side takes into account the probability that a node of class $a$ is active and reaches one of the probes, while the second term takes into account the probability that one node of class $a$ gets a connection from active probes.
Solving the integrals in Eq.~(\ref{p_a}) we can write:
\be
P_a=   \omega \left ( am+\langle a \rangle \right).
\ee
 The total number of immunized individuals, after one time step, will be:
\be
R =\sum_{a}N_aP_a=  \langle a \rangle N\omega (m+1).
\ee
The probability for having immunization for one node in activity class $a$ after $t$ time steps will be  
\be
P_a^t=1-(1-P_a)^t.
\ee
Summing over all the activity classes we can estimate the total number of immunized individuals as: 
\be
\label{R_T}
R^T =\sum_{a}N_aP_a^T=\sum_{a}N_a \left [ 1-(1-P_a)^T \right ].
\ee

The equation for $P_a$ does not consider the depletion of nodes in each class due to the immunization process. The formulation is then a good approximation for small $\omega$ and $T$ when the probability that a probe is selected more than one time is small. In this limit we observe the system for $T$ time steps and we choose the probes that have at least one connection in this time window. For each probe one of its random neighbor is immunized/removed. In this setting we study the diffusion of a SIS process. The number of infected individual in each class $a$ at the time $t+1$ can be as:
\begin{eqnarray}
\label{im_R_T}
I^{t+1}_{a} &=& I^{t}_{a}   -\mu   I^{t}_{a} + \\ \nonumber
&+& \lambda  (N_{a}-I_{a}^{t}-R_{a}^{T})a  \int d  a'  \frac{I_{a'}^{t} m }{N}+ \\ \nonumber
&+&\lambda  (N_{a}-I_{a}^{t}-R_{a}^{T})\int d  a'  \frac{I_{a'}^{t}a'  m }{N},
\end{eqnarray}
where $N_a$ is the total number of individuals with activity $a$ and $N_{a}-I_{a}^{t}-R_{a}^{T}$ is the number of susceptible nodes in the same class after $I_a^t$ moved to the class $I$ and considering that $R_a^T$ of them have been immunized/removed. Following the same lines of the previous cases we can write the threshold of this process $\xi^{ESS}$:
\be
\label{thre_NS}
\frac{\beta}{\mu} \ge \xi^{ESS} \equiv \frac{2\av{a}}{\Psi_1^T +\sqrt{ \Psi_0^T \Psi_2^T}},
\ee
where we define:
\be
\label{psi_class}
\Psi_{n}^T=\int da \,\ a^{n}(1-P_a)^TF(a).
\ee
This integral is a function of the observing time window $T$, the probability of immunization/removal of each class and the distribution of activity. In general it is not possible to solve them analytically. We evaluate  each $\Psi$ term through numerical integration of the Eq.~\ref{psi_class}. 

The threshold is a function of the moments of the activity distribution rescaled by the term $(1-P_a)^T$ which represents the fraction of nodes that are not immunized/removed in each class after an observing window $T$. These are the nodes that sustain the spreading and define the dynamical properties of the process. 

\begin{figure}[h]
\begin{center}
\includegraphics*[width=0.5\textwidth] {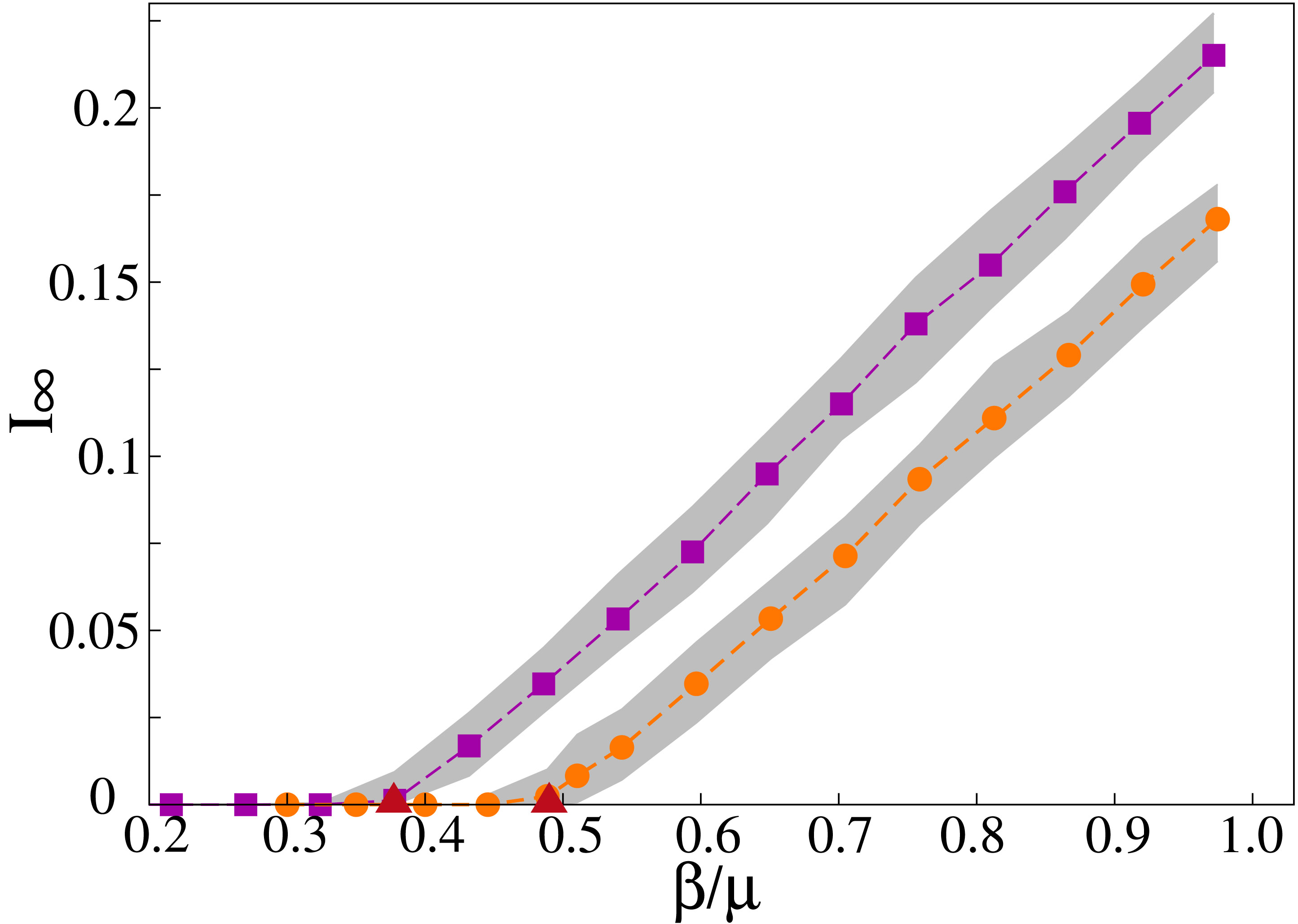}
\caption{ The plot shows the asymptotic density of a $SIS$ model, $I_\infty$, as a function of $\beta/\mu$. We consider $N=10^5$ nodes, $m=3$, $\epsilon=10^{-3}$, and a power law distribution of activity $F(a)\sim a^{-\gamma}$ with $\gamma=2.2$. In purple squares we plot the median of $I_\infty$ when no intervention is implemented. In orange circles we plot the median of $I_\infty$ when the egocentric sampling strategy of intervention is implemented. We fixed $p=0.05$ and $T=10$. The red up triangles represent the analytical prediction of the thresholds according to Eq.~\ref{thre_SIS} and Eq.~\ref{thre_NS}. In grey are plotted the $95\%$ reference range. The plot is made averaging $10^2$ independent simulations and each one of them started with $1 \%$ of random seeds.}
\label{sis_ns}
\end{center}
\end{figure}

In Figure~\ref{sis_ns} we compare the analytical results with the simulations. Considering $T=10$ and $p=0.05$ we plot the asymptotic value of the density of nodes in the class $I$ as a function of $\beta/\mu$ when no interventions are applied (purple squares) and when the egocentric strategy is implemented (orange circles). The red up triangles represent the analytical prediction of the thresholds according to Eq.~\ref{thre_SIS} and Eq.~\ref{thre_NS} that nicely reproduce the simulations. It is important stressing that the fraction of immunized/removal nodes for $T=10$ is smaller than $5\%$ since $T<T^*$ in this case.\\

\end{document}